\newcommand \Mpc {h^{-1}{\rm Mpc}}

\newcommand \kms {{\rm km~s}^{-1}}
\newcommand \msun {h^{-1} M_\odot}
\newcommand \beqn {\begin{equation}}
\newcommand \eeqn {\end{equation}}

\documentclass[apjl]{emulateapj}

\begin{document}

\title{WMAP5 and the Cluster Mass Function}

\author{Kenneth Rines\altaffilmark{1},
Antonaldo Diaferio\altaffilmark{2,3}, 
and Priyamvada Natarajan\altaffilmark{4}} 
\email{krines@astro.yale.edu}

\altaffiltext{1}{Smithsonian Astrophysical Observatory, 60 Garden St, MS 20, Cambridge, MA 02138; krines@cfa.harvard.edu}
\altaffiltext{2}{Universit\`a degli Studi di Torino,
Dipartimento di Fisica Generale ``Amedeo Avogadro'', Torino, Italy; diaferio@ph.unito.it}
\altaffiltext{3}{Istituto Nazionale di Fisica Nucleare (INFN),
Sezione di Torino, Via P.~Giuria 1, I-10125, Torino, Italy}
\altaffiltext{4}{Yale Center for Astronomy and Astrophysics, Yale University, PO Box 208121, New Haven, CT 06520-8121; priya@astro.yale.edu}

\begin{abstract}

The recently revised cosmological constraints from the Five-Year WMAP
data ameliorate previous tension between cosmological constraints from
the microwave background and from cluster abundances.  We demonstrate
that the revised estimates of cosmological parameters are in excellent
agreement with the mass function of X-ray clusters in the Sloan
Digital Sky Survey.  Velocity segregation between galaxies and the
underlying dark matter could cause virial mass estimates to be biased,
causing the mass scale of the mass function to be offset from the true
value.  Modest velocity segregation
($\sigma_{gxy}/\sigma_{DM}$=1.13$^{+0.06}_{-0.05}$) is sufficient to
match the mass function to the Five-Year WMAP results.  When the new
WMAP results are combined with constraints from supernovae and baryon
acoustic oscillations, there is no need for velocity segregation
($\sigma_{gxy}/\sigma_{DM}$=1.05$\pm$0.05).  This result agrees with
expectations for velocity segregation from state-of-the-art numerical
simulations of clusters.  Together with the improved agreement between
the new WMAP results and recent cosmic shear measurements, this result
demonstrates that the amplitude of large-scale structure in the nearby
universe matches that predicted from the structure seen in the
microwave background.  The new constraint we place on velocity
segregation in clusters indicates that virial mass estimates for
clusters are reasonably accurate.  This result suggests that future
cluster surveys will be able to probe both cosmological parameters and
fundamental cluster physics.

\end{abstract}


\section{Introduction}

The increased statistical power and sophistication of different
cosmological probes enables new tests of longstanding astrophysical
problems.  We compare recent cosmological constraints from
observations of the microwave background with those obtained from
cluster abundance estimates.  This analysis effectively compares the
amplitude of large-scale structure at $z$$\approx$1100 with the
large-scale structure in the local universe.

Clusters of galaxies are the most massive gravitationally relaxed
systems in the universe, so the observed cluster mass function is a
sensitive probe of cosmological parameters.  Galaxy cluster abundances
are most sensitive to the matter density of the universe $\Omega_m$
and $\sigma_8$, the rms fluctuations in spheres of radius 8$\Mpc$ and
the normalization of the linear power spectrum
\citep{henry91,bahcall93}.  The estimated values of these two
parameters 
underwent significant revisions between the One-Year and Three-Year
WMAP results \citep[hereafter WMAP1 and WMAP3;][]{spergel06}.  With
the recent release of the Five-Year WMAP results (hereafter WMAP5),
the estimates of these two parameters shifted again, although this
latest shift is smaller than the statistical uncertainties
\citep[][see their Table 2 and Figure 6]{dunkley08}.

The shift in these parameters between WMAP1 and WMAP3 significantly
alters the expected abundance of massive clusters.  \citet{reiprich06}
noted that this revision agrees well with the X-ray mass function of
\citet{hiflugcs}.  However, other studies suggest that hydrostatic
X-ray mass estimates such as those used in \citet{hiflugcs}
underestimate true cluster masses because either the gas
temperatures are underestimated \citep{rasia05} or because turbulent
flows and kinetic pressure in the intracluster medium (ICM) are a
significant energy component \citep{vikhlinin06,nagai07}.  The
(poorly constrained) amount of energy in non-thermal pressure has a
dramatic impact on estimates of $\sigma_8$ inferred from X-ray
observations \citep[e.g., see Figure 17 of][]{mantz08}.

We recently estimated the cluster mass function using ROSAT X-ray
cluster surveys to determine the selection function and redshifts from
the Sloan Digital Sky Survey \citep[SDSS,][]{sdss} to estimate virial
masses \citep{cirsmf}.  The cluster masses were computed as part of
the Cluster Infall Regions in SDSS project \citep[CIRS,][]{cirsi}.
Our estimate of cluster abundance exceeded that expected for the
best-fit WMAP3 parameters.  We showed that this difference could be
attributed to velocity segregation: if cluster galaxies have a
significantly higher velocity dispersion than the underlying dark
matter, then dynamical mass estimates based on the virial theorem
overestimate cluster masses and thus overestimate the abundance of
clusters at a given mass threshold.  We argued that significant
velocity segregation ($\sigma_{gxy}/\sigma_{DM}$$\approx$1.28$\pm$0.06)
is needed to produce agreement with the cosmological parameters of
WMAP3 \citep{cirsmf}.

Numerical simulations including models of galaxy formation typically
predict little velocity segregation (also called ``velocity bias'') 
in clusters
\citep{kauffmann1999a,kauffmann1999b,gao04,diemand04,faltenbacher05,faltenbacher06,biviano06,benatov06,evrard07}.
\citet{evrard07} showed that velocity dispersions of dark matter
halos provide robust and accurate mass estimates over many orders of
magnitude in halo mass; that is, dark matter particles obey virial
scaling relations.  Further, they confirmed that the CIRS velocity
dispersion function requires significant velocity segregation in
clusters to match the cosmological parameters estimated from a joint
analysis of WMAP3 and SDSS data \citep{tegmark07}.  In particular,
they show that the parameter $S_8=\sigma_8(\Omega_m/0.3)^{0.35}$ has
conflicting estimates of $S_8$=0.69 from WMAP3+SDSS versus
$S_8$$\approx$0.9 from the CIRS velocity dispersion function.
\citet{evrard07} also show that the WMAP3+SDSS cosmology requires
excess specific energy in the ICM, again conflicting with expectations
from numerical simulations.  They propose that a plausible
intermediate value of $S_8$=0.8 would reconcile the estimates without
requiring excess specific energy in galaxies or the ICM.

In this Letter, we show that the revised cosmological constraints from
WMAP5 and other methods significantly alleviate the need for velocity
segregation and excess ICM specific energy in galaxy clusters.  We
review the challenges of accurately determining the cluster mass
function in $\S 2$ and compare the WMAP5 results with the cluster mass
function in $\S 3$.  We discuss the implications of these comparisons
in $\S 4$.  We assume $H_0 = 100 h~\kms \mbox{Mpc}^{-1}$, and a flat
$\Lambda$CDM cosmology ($\Omega _\Lambda = 1-\Omega _m$) throughout.
Where not stated explicitly, we assume $\Omega_m=0.3$ and $h$=0.7 for
initial calculations.

\section{Determining the Cluster Mass Function}

There are many challenges to accurately estimating the abundance of
galaxy clusters.  The first challenge is accurately measuring the
survey volume.  In particular, one must know the maximum volume
$V_{max}$ in which a cluster with certain properties could be detected
within a given survey.  Most clusters exhibit X-ray emission from the
ICM. As a result, X-ray surveys can be used to define flux-limited
samples where $V_{max}$ can be computed directly.  Optical surveys are
excellent for detecting clusters, but they are more sensitive to
projection effects than X-ray surveys.  Our approach was therefore to
combine the virtues of a well-defined selection function available
from large-area X-ray surveys with the large redshift samples
available from SDSS \citep{cirsmf}.  This approach has the advantage
that cluster selection and cluster mass estimates are decoupled.

Another significant challenge is sampling a large and representative
volume.  Clusters are extremely rare objects, so a large survey volume
is required to fairly sample the cluster mass function.  
The cluster sample we use is the CIRS
sample of \citet{cirsi} with some minor modifications described in
\citet{cirsmf}.  The CIRS clusters are selected from X-ray cluster
catalogs constructed from ROSAT All-Sky Survey data \citep{ebcs,noras,2001A&A...369..826B}.  
We searched for all $z$$\leq$0.1 X-ray clusters within the
4783 square degrees of spectroscopic data available in the Fourth Data
Release (DR4) of SDSS \citep[][]{dr4}.  
The volume contained in the CIRS mass function is $\sim$10$^7
h^3$Mpc$^{-3}$, the largest ever probed by virial mass estimates
\citep{cirsmf}.  

The CIRS clusters are an unbiased sample: the selection of the sample
is based purely on X-ray flux and the footprint of the SDSS DR4
spectroscopic survey.  We confirmed this claim with a $V/V_{max}$ test
\citep{schmidt68}: we find $<V/V_{max}>$=0.518$\pm$0.035 compared to
an expected value of 0.5 for a complete, uniform sample
\citep{cirsmf}.  The CIRS sample is essentially volume-limited for
clusters with $L_X$$>$3$\times 10^{43} h^{-2}$erg s$^{-1}$ and
$z$$\leq$0.1.


The greatest obstacle to measuring the cluster mass function is
obtaining sufficiently accurate mass estimates.  It is observationally
challenging to obtain detailed mass estimates for large samples of
clusters.  One common approach to computing the cluster mass function
is therefore to use relatively simple observables such as X-ray
temperature or luminosity \citep[e.g.,][]{mantz08} or optical
luminosity or richness \citep[e.g.,][]{bahcall03a,rozo08} as proxies
for detailed mass estimates.  Because these mass proxies require
significantly less data than detailed mass estimates, they can be
applied to large samples of survey-quality data and thus probe larger
volumes than CIRS.  However, an important assumption in this method is
that residuals from the scaling relation must be well-behaved
\citep[the residuals are usually assumed to follow a log-normal
distribution, e.g.,][]{mantz08}.

For the CIRS mass function, we use detailed dynamical mass estimates
for the individual clusters to compute the mass function.  The
dynamics of cluster galaxies provided the first evidence for dark
matter when \citet{zwicky1933,zwicky1937} applied the virial theorem
to the Coma cluster.  Since then, the use of virial mass estimators
has been refined and this method has proved to be a powerful tool for
measuring cluster masses \citep[e.g.,][]{biviano06}.  \citet{cirsi}
show the infall patterns of the 72 CIRS clusters and compute
the mass profiles from both the caustic technique and the virial
theorem.  With the dense redshift samples available from SDSS, cluster
members can be cleanly separated from foreground and background
galaxies, and the statistical uncertainties in the virial mass
estimates are relatively small.

Compared to X-ray studies, virial mass estimates are sensitive to
larger scales ($r_{200}$ rather than $r_{500}$, where $r_{\Delta}$ is
the radius within which the enclosed density is $\Delta$ times the
critical density).  This difference allows comparisons with
theoretical mass functions with significantly less extrapolation
\citep{white02}.  Also, virial masses can be estimated for poor
clusters and rich groups, whereas X-ray mass estimates of these
systems are complicated by possible energy input from supernovae and
AGN \citep[e.g.,][]{loewenstein00}.  Probing these smaller masses enables
a direct constraint on fluctuations on the scale 8$\Mpc$, rather than
the $\sim$14$\Mpc$ scale probed by $\sim$10$^{15}\msun$ clusters
\citep{pierpaoli01}.

We used the virial mass function to obtain cosmological constraints;
these constraints can be parameterized as
$\sigma_8(\Omega_m/0.3)^{0.5}=0.81^{+0.08}_{-0.05}$ \citep[68\%
confidence level;][]{cirsmf}.  \citet{evrard07} showed that the CIRS
velocity dispersion function yields similar results to the CIRS mass
function.  Using stacked clusters from the maxBCG sample to determine
scaling relations, \citet{becker07} found a velocity dispersion
function in good agreement with that found by CIRS.  These comparisons
support our claim that the CIRS mass function is accurate.

%
%

\section{Comparing to WMAP5 and WMAP5+SN+BAO}

\begin{figure}
\figurenum{1}
\plotone{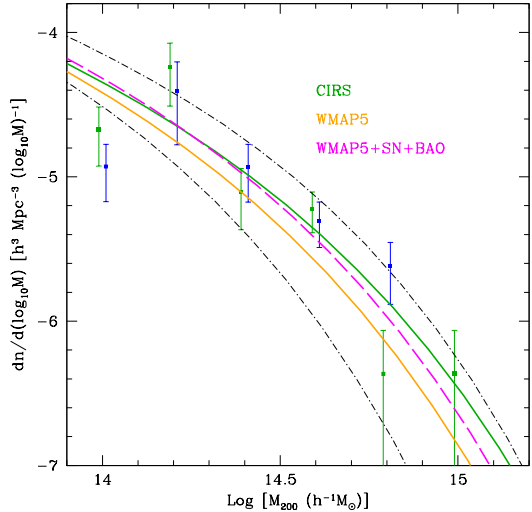}
\caption{\label{dmfn} 
The mass function of the CIRS sample.  Green and blue points and
errorbars are computed using virial masses and caustic masses
respectively (the errorbars show 68\% uncertainties).  The dash-dotted lines show the mass functions
computed using the cosmological parameters from the WMAP1 results
(upper) and WMAP3 results (lower) using the results of
\citet{jenkins01}.  The red line shows the best-fit mass
function for the CIRS virial mass function.  The orange (solid) and
 magenta (dashed) lines show the expected mass functions from the WMAP5 and
WMAP5+SN+BAO parameters respectively.}
\end{figure}

Figure \ref{dmfn} shows the mass functions for the best-fit
cosmological parameters from WMAP1 \citep{spergel03}, WMAP3
\citep{spergel06}, WMAP5 \citep{dunkley08}, and WMAP5+SN+BAO
\citep{komatsu08} at $z$=0.062, the mean redshift of the CIRS sample,
using the mass function of \citet{jenkins01}.  The WMAP5+SN+BAO
constraints combine the WMAP5 data with supernova data from recent
surveys \citep{astier06,riess07,woodvasey07} and estimates of baryon
acoustic oscillations from \citet{percival07}.
The WMAP1 and WMAP3 predictions straddle the CIRS results, while the
WMAP5 and WMAP5+SN+BAO show excellent agreement with the CIRS mass
function.  In fact, the mass function predicted by the WMAP5+SN+BAO
parameters is almost indistinguishable from the best-fit CIRS mass
function (green curve in Figure 1).

\begin{figure}
\figurenum{2}
\plotone{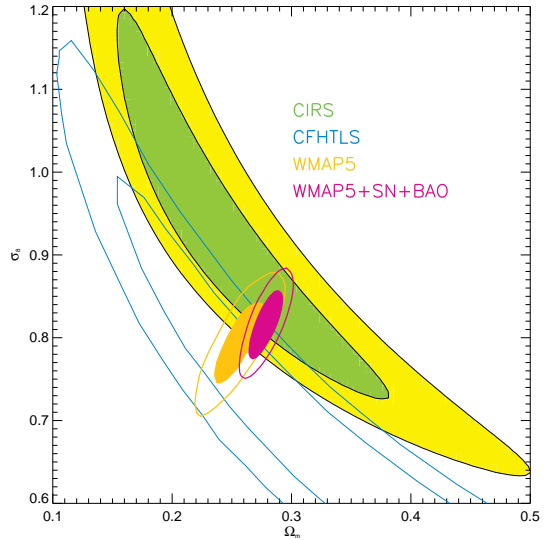}
\caption{\label{omsig} 
Cosmological constraints from the CIRS virial mass function compared
to other results.  Green/yellow contours show 68\% and 95\% confidence
levels for $\Omega_m$ and $\sigma_8$.  Blue, orange, and magenta
contours show the 68\% and 95\% confidence levels for CFHTLS, WMAP5,
and WMAP5+SN+BAO respectively
\citep[][]{fu08,dunkley08,komatsu08}.  }
\end{figure}

Figure \ref{omsig} compares the CIRS cosmological constraints in the
$(\Omega_m,\sigma_8)$ plane with constraints from WMAP5, WMAP5+SN+BAO,
and cosmic shear.  The green and yellow contours in Figure \ref{omsig}
show the 68\% and 95\% confidence levels for $\Omega_m$ and $\sigma_8$
from the CIRS virial mass function.  Blue, orange, and magenta
contours show the 68\% and 95\% confidence levels from CFHTLS
\citep[][]{fu08}, WMAP5, and WMAP5+SN+BAO respectively.
The 68\% confidence levels of WMAP3 and CIRS overlapped each other
\citep{cirsmf}, but the best-fit value of WMAP3 suggested a smaller
cluster abundance (Figure \ref{dmfn}).  Both the WMAP5 and
WMAP5+SN+BAO constraints have much smaller uncertainties than WMAP3.
The CIRS constraints are in excellent agreement with these new,
tighter constraints.  Note that cosmic shear estimates of $\Omega_m$
and $\sigma_8$ have comparable statistical uncertainties with the CIRS
constraints.  The cosmic shear constraints vary by approximately the
size of the 95\% contours when either different measures of cosmic
shear are used \citep[e.g., two-point correlation function versus
aperture mass,][]{fu08} or when applied to different samples
\citep{massey07}.  These variations suggest that the uncertainties in
cosmic shear constraints are limited by systematic effects at present.
The range of the cosmic shear constraints spans the range of the CIRS
constraints, indicating broad agreement between these very different
measures of large-scale structure.

\epsscale{1}

\section{Discussion}

We compare the recent revisions to cosmological parameter estimates
available from WMAP Five-Year data to the mass function of galaxy
clusters in the local universe.  The differences between WMAP3 and
WMAP5 have a large effect on the expected mass function.  In
particular, our determination of the cluster mass function found a
larger cluster abundance than expected for the parameters estimated by
WMAP3 \citep{cirsmf}.  The revised parameters bring the expected mass
function into much better agreement with our data.

Large velocity segregation between galaxies and the underlying dark
matter could produce a significant offset between the observed and
true mass function (due to overestimating or underestimating cluster
masses).  Such large velocity segregation is not expected from
state-of-the-art numerical simulations
\citep{diemand04,gao04,faltenbacher06,biviano06}; \citet{evrard07}
summarize the theoretical expectation as
$\sigma_{gxy}/\sigma_{DM}$=1.00$\pm$0.05.  With the WMAP5 parameters,
modest velocity segregation
($\sigma_{gxy}/\sigma_{DM}$$\approx$1.13$^{+0.06}_{-0.05}$) is
sufficient to produce agreement.  When combined with constraints from
supernovae and baryon acoustic oscillations, the implied velocity
segregation is small ($\sigma_{gxy}/\sigma_{DM}\approx$1.05$\pm$0.05),
consistent with expectations from simulations.  If the WMAP5 and CIRS
results are correct, then cluster galaxies are indeed robust tracers
of the velocity distribution (and dynamics) of the underlying dark
matter.

Similarly, the large excess specific energy required in the ICM to
match WMAP3 results with X-ray observations \citep{evrard07} is
also significantly reduced.  In the parameterization
$S_8$=$\sigma_8(\Omega_m/0.3)^{0.35}$, WMAP3 indicated $S_8$=0.69,
while a larger value of $S_8$=0.8 would imply that the ICM has
comparable specific energy to the dark matter (perhaps 10\% larger,
which is within the range expected from simulations), while the
implied ICM mass fractions would be consistent with X-ray and
Sunyaev-Zeldovich observations \citep{evrard07}.  The WMAP5 results
alone indicate $S_8$=0.75, while the WMAP5+SN+BAO results indicate
$S_8$=0.80, exactly the value suggested above as a possible resolution
of the tension between WMAP3 and X-ray cluster observations.


This agreement also supports the case for future large cluster surveys
as potential cosmological
probes\citep[e.g.,][]{haiman01,hu03b,vikhlinin03,majumdar04,mantz08}.
We conclude that a large spectroscopic program to measure virial
masses of an X-ray selected sample at moderate redshift would provide
an independent measurement of the evolution of the mass function.  If
other cosmological probes provide tighter (and consistent) constraints
on cosmological parameters than those achievable with cluster surveys,
the data can be used to probe the dynamical history of clusters, e.g.,
by measuring the evolution of velocity segregation and the thermal
history of the ICM.

\acknowledgements

We thank Adrian Jenkins for providing his software for calculating the
mass function.  We thank Margaret Geller for suggesting that we write
up this analysis.

\bibliographystyle{apj}
\bibliography{rines}

\clearpage

\epsscale{1}

\end{document}